\def\Hc{H_{\rm C}} 
\def\Hs{H_{\rm S}} 
\def\Hsb{H_{\rm Sb}}
\def\Hb{H_{\rm b}}
\def\Q{\hat Q} 
\def\A{\hat A} 
\def\tr{\mathop{\rm Tr}\nolimits} 
\def\Trb{\mathop{\rm Tr}_{\rm b}\nolimits} 
\def\taup{\tau_p}
\def\half{{1\over2}}
\def\1{\openone}
\def\F{\mathcal{F}}
\def\X{{\rm X}}
\def\Y{{\rm Y}}

\documentclass[aps,prb,tightenlines,twocolumn]{revtex4}
\usepackage{graphicx}

\def\urlprefix#1#2{\hskip0pt plus0.01fil\discretionary{}{}{}%
  {\url{#2}}}

\begin{document}

\title{Quantum kinetics of refocusing}
\author{Leonid P. Pryadko}
\affiliation{Department of Physics, University of California,
  Riverside, CA 92521} 
\author{Pinaki Sengupta}
\affiliation{Department of Physics, University of California,
  Riverside, CA 92521} 
\date{September 30, 2005}
\begin{abstract}
  We consider quantum kinetics of an open quantum system in the
  presence of periodic fields designed to suppress the internal
  evolution and shield the system from generic low-frequency
  environment (refocusing or dynamical decoupling in application to
  multi-qubit systems).  Assuming that the refocusing has order $K$,
  that is, for frozen environment the cumulant expansion of the
  evolution operator over the period $\tau$ begins with the term $\sim
  \tau^{K+1}$, we trace the associated cancellations in the kernel of
  the quantum kinetic equation in the Floquet formalism and
  characterize the remaining decoherence processes.
\end{abstract}
\maketitle
\section{Introduction}

Evolution of a quantum system subject to external time-dependent
fields is a well studied problem that goes all the way to the origins
of quantum mechanics.  However, driven dynamics in conventional atomic
physics rarely involves as intricate dynamical interference patterns
as those that occur, e.g., in multidimensional nuclear magnetic
resonance (NMR) experiments and other applications of coherent control
(CC) where precisely shaped and timed signals are used to steer the
quantum evolution of the system.  One such control
method\cite{slichter-book,hodgkinson-emsley-2000,vandersypen-2004}
originally developed in NMR is a pulse-based technique known as
dynamical recoupling (also, ``bang-bang'', in the case of hard,
$\delta$-function-like pulses).  In the simplest setup, the system is
a collection of individually-controlled weakly-coupled parts (e.g.,
qubits).  Individual qubits undergo a rapid forced precession, while
the overall long-time evolution of the system is governed by the
effective Hamiltonian averaged over their precession.  For example,
the interaction $J\hat\sigma_1^z\hat\sigma_2^z$ between the two qubits
is cancelled on average if one of them is rapidly precessing around
the $x$-axis.  Such cancellation of the quantum evolution is called
``dynamical decoupling'' or ``refocusing''; it is obviously related to
the spin-echo experiment\cite{hahn-1950}.

An exciting thing about dynamical decoupling is its universality: one
does not need to know the magnitude of the interaction precisely to
cancel it.  Moreover, with sufficiently fast pulse rate one can also
cancel the evolution due to slowly-varying external perturbations, in
effect suppressing the decoherence caused by the
environment\cite{viola-lloyd-1998,%
  vitali-tombesi-1999,%
  viola-knill-lloyd-1999,viola-knill-lloyd-1999B,%
  viola-knill-lloyd-2000,%
  vitali-tombesi-2002,%
  uchiyama-aihara-2002,%
  byrd-lidar-2003,%
  shiokawa-lidar-2004,%
  faoro-viola-2004,facchi-nakazato-2004}.  General analysis of such
methods was limited to numerics and/or the idealized
$\delta$-function-like ``hard'' pulses whose duration was either
ignored or assumed small.  In real experiments (especially in
solid-state systems), very short pulses are impractical because they tend
to couple to degrees of freedom in wide spectral range which leads to
signal distortions and heating\cite{sengupta-pryadko-ref-2005}.
Besides, from the previous studies it 
is hard to judge whether high-order refocusing sequences involving
more intricate cancellation (dynamical destructive interference) of
the quantum dynamics have a real advantage in suppressing the
decoherence.

The goal of this work is to construct a general theory of quantum
kinetics of open systems in the presence of periodic refocusing fields
designed to suppress the internal evolution and decouple the system
from outside degrees of freedom.  We describe the evolution of the
density matrix of such systems, with weak internal and thermal bath
couplings, in the approximation of master (quantum kinetic) equation
in the Floquet formalism.  The kinetics of the system is treated in a
non-Markovian
approximation\cite{konstantinov-perel-1960,%
  davies-1974,dykman-1978,alicki-1989}, which is essential to trace
the decoherence suppression with the bath ``slow'' on the scale of the
driven dynamics.

Our analysis begins with the assumption that the period-$\tau$ control
fields provide order-$K$ refocusing for the system with the bath frozen.
That is, if we replace the operators for the external degrees of
freedom by $c$-numbers, the cumulant expansion of the evolution
operator in powers of thus modified internal hamiltonian begins with
the terms of order $\sim \tau^{K+1}$.  The control fields are assumed
to be stong and we treat them exactly.  We trace the cancellations
associated with refocusing in the kernel of the quantum kinetic
equation (QKE) which describes the dissipative dynamics of the density
matrix of the system with bath present, and characterize the remaining
decoherence processes.  We illustrate the general analytic results
derived for orders $K\le 2$, which is the accuracy of the employed
QKE, with the numerical simulations of dynamics of a single spin in
the presence of a classical fluctuating magnetic field.

Our results can be summarized as follows.  Generally, for a
weakly-coupled system weakly interacting with slow degrees of freedom
(thermal bath) with the correlation time $\tau_0$, the decoherence is
due to dissipative processes (resonant decay) which create
excitation(s) in the environment, as well as reactive processes which
result in dephasing, or scrambling of the phase of the system.  The
associated decoherence rate is proportional to the square of the
coupling matrix element and the correlation time $\tau_0$, see
Eq.~(\ref{eq:decoherence-rate-stationary}).  As a result of forced
precession of the system caused by the control fields, the effective
environment seen by the system is modulated out of resonance, which
may entirely suppress the state
decay\cite{kofman-kurizki-2001,kofman-kurizki-2004}.  Only first-order
($K=1$) refocusing is necessary to achieve such an effect.  We show
that, in addition, the rate of reactive processes is reduced by a
factor of $\tau/\tau_0$ [Eq.~(\ref{eq:decoherence-rate-one})], where
the period of the refocusing sequence, $\tau$, is assumed to be
smaller than $\tau_0$.  With second-order refocusing, $K=2$, the
decoherence rate is additionally suppressed
[Eq.~(\ref{eq:decoherence-rate-two})], and with time-reversal
invariant bath coupling may even become exponentially small in this
parameter (in which case it will be determined by terms of higher
order in bath coupling, beyound the accuracy of our QKE).

In addition to the decoherence rates which characterize the
exponential decay of quantum correlations with time, we also analyze
the corresponding prefactor, which determines the initial
decoherence\cite{falci-2005}.  While for generic refocusing sequences
with $K\ge1$ initial decoherence is quadratic in $\tau$ and does not
scale with the thermal bath correlation time $\tau_0$, we show that
for symmetric pulse sequences it is reduced by an additional power of
$(\tau/\tau_0)$.

Our results extend the well-established theory\cite{dykman-1979} of
the kinetics of few-level systems in r.f.\ field to cases involving a
more intricate cancellation (dynamical destructive interference) of
the quantum dynamics characteristic of higher-order refocusing
sequences.  They put in a perspective the previous analyses of
decoherence in the
presence of hard-pulse sequences\cite{viola-lloyd-1998,%
  vitali-tombesi-1999,%
  viola-knill-lloyd-1999,viola-knill-lloyd-1999B,%
  viola-knill-lloyd-2000,%
  vitali-tombesi-2002,%
  uchiyama-aihara-2002,%
  byrd-lidar-2003,%
  shiokawa-lidar-2004,%
  faoro-viola-2004,facchi-nakazato-2004}, establish a firm basis for
future studies of decoherence scaling in large driven qubit systems
(with and without long-range coupling due to long-wavelength phonons
which may be correlated\cite{Ischi-Hilke-Dube-2005} between distant
qubits, contrary to a common assumption in the quantum
error-correction theory\cite{shor-error-correct}), and present an
efficient alternative to optimum control techniques based directly on
the master equation\cite{rabitz-science-2000-control,%
  levis-science-2001-control,ohtsuki-2003-control}. 

Some of the results regarding decoherence suppression in the presence
of higher-order refocusing sequences were announced
previously\cite{sengupta-pryadko-ref-2005}, as  a justification
for developing a technique for designing higher-order pulses and pulse
sequences.

\section{Problem setup}
{\bf Hamiltonian of the system.}
We consider $N$-level open system with the Hamiltonian
\begin{equation}
H=\Hc(t)+\Hs+\Hsb+ \Hb,\label{eq:ham-total}
\end{equation}
where the oscillator bath Hamiltonian $\Hb=\sum_\mu \omega_\mu
a_\mu^\dagger a_\mu$ has the usual form, while the control Hamiltonian
\begin{equation}
  \label{eq:ham-control}
  \Hc\equiv\half\sum_\alpha V_\alpha(t) \Sigma_\alpha,
\end{equation}
the system Hamiltonian 
\begin{equation}
  \label{eq:ham-system}
  \Hs\equiv\half\sum_\alpha J_\alpha \Sigma_\alpha,
\end{equation}
and the system-bath coupling Hamiltonian
\begin{equation}
  \label{eq:ham-system-bath}
  \Hsb\equiv\half\sum_\alpha  b_\alpha \Sigma_\alpha, \quad 
  b_\alpha=\sum_\mu {f_{a\mu} a_\mu+f_{a\mu}^* a_\mu^\dagger\over (2m_\mu
  \omega_\mu)^{1/2}},
\end{equation}
are expressed in terms of $N\times N$ Hermitian matrices
$\Sigma_\alpha$, $\alpha=0, \ldots, N^2-1$, normalized so that
$\tr(\Sigma_\alpha \Sigma_\beta)=N\delta_{\alpha\beta}$.  It is
convenient to specify explicitly $\Sigma_0=\1$, choose the remaining
matrices traceless, and define the algebra via the commutators and the
anticommutators,
\begin{equation}
  \bigl[\Sigma_\alpha,\Sigma_\beta\bigr]
  =2i C_{\alpha\beta}^\gamma\Sigma_\gamma,  
  \quad  
  \bigl\{\Sigma_\alpha,\Sigma_\beta\bigr\}
  =2B_{\alpha\beta}^\gamma\Sigma_\gamma.
\label{eq:algebra}
\end{equation}
For example, for a single qubit (spin), $N=2$, we can choose
$\Sigma_\alpha\equiv \sigma_\alpha$, $\alpha=0,\ldots 3$, in terms of
the unit matrix $\sigma_0\equiv1$ and the three Pauli matrices, in
which case the net coefficients
$B_\nu(t)=V_\nu(t)+J_\nu+b_\nu$, $\nu=1,2,3$, can be
interpreted as the components of the time-dependent magnetic field
acting on the spin.  

Similarly, for two-qubit system the full set can be chosen in terms of
the  direct products $\sigma_i\otimes\sigma_j$,
$i,j=0,1,2,3$.  In this case, the coefficients in front of the
single-spin operators $\sigma_{1\nu}\equiv\sigma_\nu\otimes \sigma_0$
and $\sigma_{2\nu}\equiv\sigma_0\otimes \sigma_\nu$ can be interpreted
as the components of the magnetic fields acting on the corresponding
spin, while the two-spin operators $\sigma_\nu\otimes \sigma_\rho$
describe spin couplings.

The same matrices will be used to parametrize the density matrix of
the system,
\begin{equation}
  \label{eq:density-matrix}
  \rho={1\over N}(\Sigma_0+\sum_{\alpha\ge1}R_\alpha \Sigma_\alpha).
\end{equation}
The normalization is chosen so that $\tr\rho=1$.  Also, for a pure
state, $\rho^2=\rho$, we have $R^2\equiv (R_\alpha)^2=N-1$ (summation
implicit), while for fully mixed state $R^2=0$.

Generally, only few of all $N^2$ allowed terms are expected to be
present in the Hamiltonian.  Particularly, for $n$-qubit system with
$N=2^n$, it is common to have single-qubit controlling fields
$V_{l\nu}$, $l=1,\ldots, n$, while the couplings (both intrinsic ones,
$J_\alpha$, and bath, $b_\alpha$) can be one-, two-, or multi-particle
as physically appropriate.

{\bf Dynamical decoupling in a closed system.}  Here we consider an
auxiliary control problem for the system~(\ref{eq:ham-total}) with the
thermal bath operators $b_\alpha$ in Eq.~(\ref{eq:ham-system-bath})
replaced by constant numbers, which in effect modifies the
coefficients in the system Hamiltonian~(\ref{eq:ham-system}).  The
control goal is to suppress the unitary evolution with thus modified
hamiltonian $\Hs$ as fully as possible.
Unless the fields $V_\alpha(t)$ are specified to exactly compensate the
Hamiltonian $\Hs$ (which is never practical), the refocusing can only
be achieved at some discrete set of time moments
$t_0=0$, $t_1=\tau$, \ldots.  The unitary evolution over the
refocusing interval $\tau$ is commonly analyzed in terms of the
effective Hamiltonian theory, a perturbative scheme based on the
cumulant (Magnus) expansion for the evolution
operator\cite{waugh-huber-haeberlen-1968,waugh-wang-huber-vold-1968}.
The expansion 
is done around the evolution in the applied controlling fields
[Hamiltonian $\Hc(t)$], while the system Hamiltonian $\Hs$ is treated
perturbatively.  Obviously, this implies that the controlling
Hamiltonian dominates the evolution.

Explicitly, consider the evolution operator $U(t)$,
\begin{equation}
  \label{eq:evol-oper}
  \dot U(t) =-i\, [ \Hc(t)+ \Hs]\, U(t),\quad U(0)=\1.
\end{equation}
As usual, the time-dependent perturbation theory is introduced by
separating out the bare evolution operator, 
\begin{equation}
  \label{eq:evol-bare}
  U(t)=U_0(t)\,R(t),\quad
  \dot U_0(t) =-i \Hc(t)\, U_0(t) .
\end{equation}
Then, the operator $R(t)$ obeys the equation
\begin{equation}
  \dot R(t)=-i  \Hs(t) R(t), \quad 
 \Hs(t)\equiv U_0^\dagger (t)\,   \Hs\, U_0(t),
  \label{eq:evol-oper-inter}
\end{equation}
which can be iterated to construct the standard  expansion $R(t)
  =\1+R_1(t) +R_2(t) +\ldots$ in powers of $(t\,\Hs)$, 
\begin{equation}
  \label{eq:pert-expansion}
    \dot R_{n+1}(t)=-i  \Hs(t) R_{n}(t), \quad
    R_{0}(t)=\1.
\end{equation}

The standard Magnus expansion is readily obtained by integrating
Eqs.~(\ref{eq:pert-expansion}) formally and rewriting the result in
terms of cumulants,
\begin{eqnarray}
  \label{eq:cumulants}
  R(t)&=&\exp\biglb({C_1(t)+C_2(t)+\ldots}\bigrb),\\
  \label{eq:cumulant-one}
  C_1(t)&=&-i\int_0^t dt_1  \Hs(t_1),\\ 
  \label{eq:cumulant-two}
  C_2(t)&=&-{1\over 2}\int_0^t dt_2\int_0^{t_2} dt_1
  \bigl[\Hs(t_1),\Hs(t_2)\bigr],\cdots .
\end{eqnarray}
Generally, the term $C_k$ contains a $k$-fold integration of the
commutators of the rotating-frame Hamiltonian $\Hs(t_i)$ at different
time moments $t_i$ and has an order $(t \Hs)^k$.  For a given
Hamiltonian $\Hs$, order-$K$ refocusing is characterized by vanishing
of the terms $C_k$ of order $k\le K$ at the given time moment
$t=\tau$.  This is equivalent to the condition $R_k(\tau)=0$ for
$k=1,\ldots, K$.  The latter matrices can be efficiently evaluated
numerically, which gives a systematic method for analysis and
optimization of the controlled dynamics in high orders of the
cumulant expansion\cite{sengupta-pryadko-ref-2005}.

In terms of the matrices $\Sigma_\alpha$, the unitary transformation
generated by the control fields 
amounts to a rotation, 
\begin{equation}
  U_0(t)\Sigma_\alpha U_0^\dagger(t) \equiv
  Q_{\alpha\beta}(t)\Sigma_\beta, 
\label{eq:rotation}
\end{equation}
where the matrix $\Q(t)$ is orthogonal, $\Q^{\rm tr}=\Q^{-1}$.  We
assume the matrix to be periodic with the refocusing period $\tau$,
$\Q(t)=\Q(t+\tau)$, which will be referred to as the ``zeroth order''
refocusing condition.  This condition is non-trivial; it does not
reduce to the periodicity of the control fields $V_\alpha(t)$.  

The periodicity of the real-valued matrices $\Q(t)$ implies the
Fourier expansion with the frequencies $\Omega_m\equiv 2\pi m/\tau$,
\begin{equation}
  \label{eq:rotation-fourier}
  \Q(t)=\sum\nolimits_m \A_m e^{i\Omega_m t}, \quad \hat A_{-m}=\hat A_m^*.   
\end{equation}
From orthogonality, $\Q(t) \Q^{\rm tr}(t)=\hat1$, we have 
\begin{equation}
  \sum\nolimits_k \A_k \A_{m-k}^{\rm tr}=\delta_{m,0}\,\hat1.
  \label{eq:sum-rule-A}
\end{equation}
With these definitions, it is easy to rewrite the first two refocusing
conditions in algebraic form.  Specifically, 
$$
i C_1(\tau)={1\over 2}J_\alpha \Sigma_\beta \int_0^\tau dt\,
Q_{\alpha\beta}(t),
$$
and the first-order refocusing condition, $C_1(\tau)=0$, is
\begin{equation}
  \label{eq:refocusing-one}
  [\hat A_{0}^{\rm tr}]_{\beta\alpha}J_\alpha=0 ,\;\,\textrm{or just}\;\,
 \hat A^{\rm tr}_0 J=0,
\end{equation}
where in the second form of the expression we treated the coefficients
$J_\alpha$ as a column vector.  

Performing the double integration in Eq.~(\ref{eq:cumulant-two}) in
the assumption that the first-order refocusing condition is satisfied,
we have for the second-order refocusing, $C_2(\tau)=0$,
\begin{equation}
  \label{eq:refocusing-two}
  C_{\alpha\beta}^\gamma\sum_{m\neq0}{[\hat A_{-m}^{\rm
    tr}\,J\, J^{\rm tr}\, \hat A_m]_{\alpha\beta}\over i\Omega_{-m}}=0,
\end{equation}
where the coefficients $C_{\alpha\beta}^\gamma$ define the
commutators, see Eq.~(\ref{eq:algebra}).  We note that the sum in
Eq.~(\ref{eq:refocusing-two}) is antisymmetric with respect to indices
$\alpha$, $\beta$, and an analogous condition with the symmetric
coefficients $B_{\alpha\beta}^\gamma$ [which define anticommutators in
  Eq.~(\ref{eq:algebra})] is trivially satisfied.

\section{Quantum kinetics.}
{\bf Quantum kinetic equation in rotating frame.}  In this work we
consider slow (on the scale of the refocusing period $\tau$)
environment, which makes it necessary to consider quantum dynamics of
the system outside the commonly used Markovian approximation.  We
write the master equation 
as\cite{konstantinov-perel-1960,davies-1974,dykman-1978,alicki-1989}
\begin{eqnarray}
  \label{eq:qke-general}\nonumber
  \dot \rho(t)& =& -i\bigl[\langle H_1(t)\rangle ,\rho(t)\bigr]\\
  & &\!\!\!\!\!\! -\int_0^t\!\! dt'\,\Trb
\bigl[\delta H_1(t),\bigl[\delta
  H_1(t'),\rho(t')\rho_{\rm b} \bigr]\bigr],
\end{eqnarray}
where $\langle  H_1(t)\rangle\equiv \Trb \biglb(
H_1(t)\,\rho_{\rm b}\bigrb)$ and $\delta H_1(t)\equiv 
H_1(t)-\langle H_1(t)\rangle$.  Here $ H_1(t)$ is the
interaction representation of the perturbation Hamiltonian $H_1\equiv
\Hs+\Hsb$ [see Eq.~(\ref{eq:ham-total})] in the rotating frame
generated by the control and the thermal bath parts of the
Hamiltonian, $H_0\equiv \Hc+\Hb$, and the bath is assumed to be in
thermal equilibrium, $\rho_{\rm b}\equiv \exp(-\beta\Hb)/Z$, $Z\equiv
\Trb\exp(-\beta\Hb)$.

With the definitions~(\ref{eq:ham-system}), (\ref{eq:ham-system-bath})
the average perturbation Hamiltonian in the first term of the
QKE~(\ref{eq:qke-general}) is given just by
Eq.~(\ref{eq:evol-oper-inter}),
\begin{equation}
\langle  H_1(t)\rangle=\Hs(t)=\half [\Q^{\rm tr}(t)\,
  J]_\alpha\Sigma_\alpha,\label{eq:ham-averaged-rotating}
\end{equation}
while the corresponding fluctuating part is 
\begin{equation}
\delta H_1(t)=\Hsb(t)=\half Q^{\rm tr}_{\beta\alpha}(t)
   b_\alpha(t)\,\Sigma_\beta,\label{eq:ham-fluct-rotating}
\end{equation}
where the oscillator fields in the interaction representation $
b_\alpha(t)$ are given by Eq.~(\ref{eq:ham-system-bath}) with the
replacement $a_\mu\to a_\mu e^{i\omega_\mu t}$.  

The second term in the r.h.s.\ of the QKE~(\ref{eq:qke-general}) is
evaluated in terms of two correlators,
\begin{eqnarray}
  \label{eq:phonon-correlator}
  \F_{\alpha\beta}(t-t')&=&\Trb\{b_\alpha(t)b_\beta(t')\rho_{\rm
  b}\}, 
  \\    
  \bar\F_{\alpha\beta}(t-t')&=&\Trb
  \{b_\alpha(t)\rho_{\rm b} b_\beta(t')\},
\end{eqnarray}
which in turn can be conveniently expressed in terms of the spectral
coupling matrix (function)
\begin{equation}
  \label{eq:coupling-matrix}
  F_{\alpha\beta}(\omega)\equiv {\pi\over2}\sum_\mu {f_{\alpha\mu}
  f^*_{\beta\mu}\over m\omega_\mu}\delta(\omega-\omega_\mu).
\end{equation}
The fastest response time of the environment is characterized by the
largest frequency of an oscillator present in the system.  We will
introduce the cut-off frequency $\omega_c$, such that
$F_{\alpha\beta}(\omega)$ is only non-zero for $\omega<\omega_c$.  In
addition, we will characterize the bath with a possibly slower
``correlation time'' $\tau_0$, which describes the width of typical
features of the spectral coupling function $F_{\alpha\beta}(\omega)$.

Operators $b_\alpha(t)$ are Hermitian, thus $\hat\F^\dagger(t)
=\hat\F(-t)$.  An explicit calculation gives
$\bar\F_{\alpha\beta}(t)=\F_{\alpha\beta}^*(t)$, and
\begin{equation}
\hat\F(t)=\!\!
\int_0^\infty\! {d\omega\over \pi}\bigl [ \hat F(\omega)(n_\omega+1)
  e^{i\omega t}+ \hat F^*(\omega)\,n_\omega
  e^{-i\omega t}\bigr].
\label{eq:correlator-explicit}
\end{equation}
It is convenient to split this correlator onto real and imaginary
part, $\hat\F(t)=\hat\F_1(t)+\hat\F_2(t)$.  The
QKE~(\ref{eq:qke-general}) becomes 
\begin{eqnarray}
  \lefteqn{\dot \rho(t) = - {i\over2} 
    Q^{\rm tr}_{\alpha\beta}(t)  J_\beta
    \bigl[\Sigma_\alpha,\rho(t)\bigr]}& &  
  \nonumber\\& &\!\!\!\!\!
  -{1\over4}\!\int_0^t\!\!\!
  dt'\, [\hat Q^{\rm tr}(t) \hat \F_1(t-t') \hat
    Q(t')]_{\alpha\beta}\bigl[\Sigma_\alpha,
    \bigl[\Sigma_\beta,\rho(t')\bigr]\bigr]
  \nonumber\\& &\!\!\!\!\!
  -{i\over4}\!
  \int_0^t\!\!\! dt'\, [\hat Q^{\rm tr}(t) \hat \F_2(t-t') \hat
    Q(t')]_{\alpha\beta}\bigl[\Sigma_\alpha,
    \bigl\{\Sigma_\beta,\rho(t')\bigr\}\!\bigr].\;\;\;\;\;
  \label{eq:qke-generalized}
\end{eqnarray}
This can be further simplified with the help of the
definitions~(\ref{eq:algebra}), (\ref{eq:density-matrix}):
\begin{eqnarray}
  \dot R_\gamma(t) &=  &
    Q^{\rm tr}_{\alpha\beta}(t)  J_\beta
    C_{\alpha\delta}^\gamma R_\delta(t)
    \nonumber\\& &\!\!\!\!
    +\!\int_0^t\!\!\!
    dt'\, [\hat Q^{\rm tr}(t) \hat \F_1(t-t') \hat
      Q(t')]_{\alpha\beta}
    C_{\alpha\alpha'}^\gamma
    C_{\beta\delta}^{\alpha'} R_\delta(t')
    \nonumber\\& &\!\!\!\!
    +\!
    \int_0^t\!\!\! dt'\, [\hat Q^{\rm tr}(t) \hat \F_2(t-t') \hat
      Q(t')]_{\alpha\beta} C_{\alpha\alpha'}^\gamma
    B_{\beta\delta}^{\alpha'} R_\delta(t')
    \nonumber\\
    & &\!\!\!\! +\int_0^t\!\!\! dt'\, [\hat Q^{\rm tr}(t) \hat \F_2(t-t') \hat
    Q(t')]_{\alpha\beta} C_{\alpha\beta}^\gamma,
  \label{eq:qke-commuted}
\end{eqnarray}
where the last term comes from the (time-independent) part of the
density matrix~(\ref{eq:density-matrix}) proportional to
$\Sigma_0\equiv \1$.  This term is responsible for establishing the
equilibrium at large $t$ (dynamical equilibrium with refocusing).

We note that the structure of the QKE [in particular,
  Eqs.~(\ref{eq:correlator-explicit}), (\ref{eq:qke-commuted})]
remains the same even if the nature of the thermal bath coupling is
changed (e.g., by adding non-linear oscillator couplings) as long as
the bath remains in thermal equilibrium.  In such cases, the only
change would be a renormalization of the average
Hamiltonian~(\ref{eq:ham-averaged-rotating}) and of the coupling
matrix~(\ref{eq:coupling-matrix}).

{\bf Kinetics in the absence of control.}
In the absence of refocusing, $\Q(t)=\hat 1$, the
QKE~(\ref{eq:qke-commuted}) does not depend on time explicitly, and it
can be solved with the help of the Laplace transform [denoted with
  tilde, $f(t)\to\tilde f(p)$], 
\begin{equation}
  \label{eq:qke-stationary-laplace}
  p \tilde R_\gamma(p)-R_\gamma(0)=\Pi_{\gamma\delta}(p)
  \tilde R_\delta(p)+ 
  \tilde \F_{2,\alpha\beta}(p)C_{\alpha\beta}^\gamma,
\end{equation}
where the kernel 
\begin{equation}
  \label{eq:qke-stationary-kernel}
  \Pi_{\gamma\delta}=J_\alpha C_{\alpha\delta}^\gamma+
  C_{\alpha\alpha'}^\gamma \biglb(\tilde \F_{1,\alpha\beta}(p)
  C_{\beta\delta}^{\alpha'} +\tilde \F_{2,\alpha\beta}(p)
  B_{\beta\delta}^{\alpha'} \bigrb)
\end{equation}
incorporates the first three terms in the r.h.s.\ of
Eq.~(\ref{eq:qke-commuted}).  

The dissipative dynamics of the system is defined by the singularities
of the matrix $[\hat 1\,p-\hat \Pi(p)]^{-1}$, whose location on the
complex plane $p$ determine the spectrum of the decoherence rates.
For long-time dynamics, only the singularities close to the imaginary
axis are relevant.  With both the intrinsic interactions $J$ and bath
couplings $\F$ weak on the scale of the bath correlation time
$\tau_0$, a good accuracy can be obtained by setting $p\to0$ in the
QKE kernel $\hat\Pi(p)$ [Eq.~(\ref{eq:qke-stationary-kernel})], which
is equivalent to the Markovian bath approximation.  Then, the real
parts of the eigenvalues of the matrix $\hat\Pi(0)$ will determine the
spectrum of the decoherence rates.  If we bunch together all processes
causing the evolution of the density matrix, the maximum decoherence
rate can be estimated as
\begin{equation}
  \Gamma_0\sim\max\bigl[J,\,\Delta(0)\,\tau_0\bigr],
  \label{eq:decoherence-rate-stationary}
\end{equation}
where $\Delta(t)\equiv \|\hat\F(t)\|$ is a norm of the correlator
matrix~(\ref{eq:correlator-explicit}), and $\tau_0$ is the bath
correlation time introduced below Eq.~(\ref{eq:coupling-matrix}).

{\bf Quantum kinetics in Floquet formalism.}
The full kinetic equation~(\ref{eq:qke-commuted}) in the presence of
refocusing can also be analyzed with the help of the Laplace
transformation, but in this case the structure of the solution is
complicated by the presence of the time-dependent rotation matrices
$\Q^{\rm tr}(t)$, $\Q(t')$.  Assuming these are periodic
(``zeroth-order'' refocusing condition), we use the
expansion~(\ref{eq:rotation-fourier}) to obtain [cf.\ 
  Eq.~(\ref{eq:qke-stationary-kernel})]
\begin{equation}
  \label{eq:qke-full-laplace}
  p\tilde R_\gamma(p)=
  \sum_m \Pi_{m,\gamma\delta}(p)\tilde R_\delta(p-i\Omega_m)
  + r_{\gamma}(p), 
\end{equation}
where the kernel with the frequency transfer $\Omega_m$ is 
\begin{widetext}
\begin{eqnarray}
  \label{eq:qke-full-kernel}
  \Pi_{m,\gamma\delta}(p)&=&
  ( A_m^{\rm tr})_{\alpha\beta} J_\beta C^\gamma_{\alpha\delta}
  +C_{\alpha\alpha'}^\gamma\sum_{m'}\int {d\omega\over 2\pi}\biggl\{
  {[\hat A_{m-m'}^{\rm tr}\tilde\F_1(\omega)\hat A_{m'}]_{\alpha\beta}
    \over p-i\omega -i\Omega_{m-m'}} 
  C_{\beta\delta}^{\alpha'}+
  {[\hat A_{m-m'}^{\rm tr}\tilde\F_2(\omega)\hat A_{m'}]_{\alpha\beta}
    \over p-i\omega -i\Omega_{m-m'}}
  B_{\beta\delta}^{\alpha'} 
  \biggr\}, 
\end{eqnarray}
and the 
second term in the r.h.s.\ of
Eq.~(\ref{eq:qke-full-laplace}) is 
\begin{eqnarray}
  \label{eq:qke-full-rhs}
r_{\gamma}(p)&  =& 
  R_\gamma(0)+\sum_m {s_{m,\gamma}(p)\over p-i\Omega_m},\\
s_{m,\gamma}(p)&\equiv &
\!\sum_{m'}\! \int\! {d\omega\over 2\pi}{[\hat
  A_{m-m'}^{\rm tr}\tilde\F_2(\omega) \hat A_{m'}]_{\alpha\beta}\,
  C_{\alpha\beta}^\gamma  \over p-i\omega
    -i\Omega_{m-m'}}.
  \label{eq:qke-full-rhs-coeff}
\end{eqnarray}

The obtained expression~(\ref{eq:qke-full-laplace}) is a set of
functional equations for the Laplace-transformed matrix elements
$\tilde R_\gamma(p)$ of the density matrix~(\ref{eq:density-matrix}).
We iterate these equations to obtain a formal series in powers of
$\hat\Pi_m$,
\begin{equation}
  \tilde R(p)={r(p)\over p}+\sum_{m_1}{\hat\Pi_{m_1}(p)\over p}
  {r(p_1)
    \over p_1}
  + \!\!\!\sum_{m_1,m_2}\!\!\!{\hat\Pi_{m_1}(p)\over p}
 {\hat\Pi_{m_2-m_1}(p_1)\over p_1} 
  {r(p_2)
    \over   p_2}
  + \sum_{\{m\}}{\hat\Pi_{m_1}(p)\over p}
  {\hat\Pi_{m_2-m_1}(p_1)\over p_1} 
  {\hat\Pi_{m_3-m_2}(p_2)\over p_2} 
  {r(p_3)
    \over p_3}+\cdots,\quad
  \label{eq:iterated-series}
\end{equation}
where $p_n\equiv p-i\Omega_{m_n}$.  Generally, long-time behavior
corresponds to small values of $p$, while that near the ends of the
refocusing interval is governed by values of $p$ close to any
$i\Omega_l$, resulting in an asymptotic decomposition,
$R_\alpha(t)=\sum_l R_\alpha^{[l]} e^{i\Omega_l t -\gamma_l t}$.  We
will analyze the terms with different $l$ separately, beginning with
$l=0$.  

To evaluate the dynamics around a given frequency $\Omega_l$,
we need to carefully account for the terms singular near
$p=i\Omega_l$.  To this end, denote the sum of all non-singular (for
small $p-i\Omega_l$ with given $l$) terms connecting the terms with
the denominators $p'\equiv p-i\Omega_{l'}$ and $p''\equiv
p-i\Omega_{l''}$,
\begin{equation}
  \hat \Pi^{[l]}_{l',l''}(p)\equiv
  \hat\Pi_{l''-l'}(p')+\sum_{m_1\neq l}
  \hat\Pi_{m_1-l'}(p'){\hat\Pi_{l''-m_1}(p_1)\over p_1}
  +  \sum_{m_1,m_2\neq l}
  \hat\Pi_{m_1-l'}(p')
  {\hat\Pi_{m_2-m_1}(p_1)\over p_1}
  {\hat\Pi_{l''-m_2}(p_2)\over p_2}+\cdots.
  \label{eq:kernel-non-singular}
\end{equation}
Note that this definition implies 
\begin{equation}
  \hat \Pi^{[l]}_{l',l''}(p)=\hat
  \Pi^{[0]}_{l'-l,l''-l}(p-i\Omega_l).\label{eq:kernel-non-singular-identity}
\end{equation}
The entire series near $p=i\Omega_l$, $l\neq 0$, can now be written
as
\begin{equation}
  \tilde R^{[l]}(p)
  ={r(p)\over p}+{\hat\Pi_{0,l}^{[l]}(p)\over p}
  \Bigl(p-i\Omega_l-\hat\Pi_{l,l}^{[l]}(p)\Bigr)^{-1}
  \Bigl[r(p-i\Omega_l)+\sum_{m\neq l}\hat \Pi_{l,m}^{[l]}(p)
  {r(p-i\Omega_m)\over 
    p-i\Omega_m}\Bigl],
\label{eq:series-resum-l}
\end{equation}
while near $p=0$ it is   
  \begin{equation}
  \tilde R^{[0]}(p)
  =\Bigl(p-\hat\Pi_{0,0}^{[0]}(p)\Bigr)^{-1}\Bigl[R(0)+\!\sum_{m}{s_m(p)\over
      p-i\Omega_m} +\!\!\sum_{m\neq 0}{\hat \Pi_{0,m}^{[0]}(p)\over
      p-i\Omega_m}R(0)+\!\!\sum_{m\neq0}{\hat \Pi_{0,m}^{[0]}(p)\over
      p-i\Omega_m}\sum_{m'}{s_{m'}(p-i\Omega_m)\over
      p-i\Omega_m-i\Omega_{m'}}\Bigl],
  \label{eq:series-resum-zero}
\end{equation}
\end{widetext}
where Eq.~(\ref{eq:qke-full-rhs}) for $r(p)$ was substituted for
completeness.

The analysis of the obtained expressions is dramatically simplified
with at least order-one refocusing, as long as the couplings
$J_\alpha$ and the bath couplings $\hat\F(t)$ are weak on the
refocusing time scale $\tau$, which is assumed to be short on the
scale of the bath correlation time $\tau_0$, $\tau\ll\tau_0$.

Indeed, the universal {\em order-one refocusing\/} condition,
Eq.~(\ref{eq:refocusing-one}), implies the disappearance of the
average Hamiltonian regardless of the specific values of the couplings
$J_\alpha$.  Thus, in the kernel $\hat\Pi_m(p)$
[Eq.~(\ref{eq:qke-full-kernel})] with $m=0$, the first term disappears
completely.  Furthermore, if we assume the set of the fluctuating
fields $b_\alpha$ in the bath coupling
hamiltonian~(\ref{eq:ham-system-bath}) is the same as that of the
constant parameters $J_\alpha$ in the system
hamiltonian~(\ref{eq:ham-system}), the terms with resonant denominator
($m-m'=0$) inside the $\omega$ integrals in
Eqs.~(\ref{eq:qke-full-kernel}), (\ref{eq:qke-full-rhs-coeff}) will
also be suppressed.  Then, for small $|p|$, the resonant contribution
to these expressions will be limited by $|\omega|\ge \Omega$, which by
assumption is far out in the tail region of the spectral coupling
function~(\ref{eq:coupling-matrix}).  The remaining non-resonant
contributions can be calculated by expanding in powers of $\omega$
under the integrals.

In particular, the spectrum of the dissipation rates is determined by
the positions of the singularities of the QKE resolvent,
$\biglb(p-\hat\Pi_{0,0}^{[0]}(p)\bigrb)^{-1}$ [see
  Eq.~(\ref{eq:series-resum-zero})].  At small coupling these are
determined by the eigenvalues of the matrix $\hat\Pi_{0,0}^{[0]}(p=0)$
[Eq.~(\ref{eq:kernel-non-singular})].  To quadratic order in powers of
perturbing Hamiltonian, with the help of
Eq.~(\ref{eq:qke-full-kernel}), we have
\begin{widetext}
  \begin{equation}
    \label{eq:kernel-non-singular-expanded}
    \Bigl[\hat\Pi_{0,0}^{[0]}(p)\Bigr]_{\gamma\delta}=\!
    \sum_{m\neq0}
    {[\hat A_{-m}^{\rm tr} J \,J^{\rm tr} \hat
    A_{m}]_{\alpha\beta} 
      \over p-i\Omega_{-m}} 
  C_{\alpha\alpha'}^\gamma C_{\beta\delta}^{\alpha'}+
  \sum_{m\neq0}\sum_{k\ge 0}
    C_{\alpha\alpha'}^\gamma \biggl\{
  {[\hat A_{-m}^{\rm tr} \hat\F_1^{(k)}(0) \hat
      A_{m}]_{\alpha\beta} 
    \over (p-i\Omega_{-m})^{k+1}} 
 C_{\beta\delta}^{\alpha'}
+  {[\hat A_{-m}^{\rm tr} \hat\F_2^{(k)}(0) \hat
      A_{m}]_{\alpha\beta} 
    \over (p-i\Omega_{-m})^{k+1}} 
 B_{\beta\delta}^{\alpha'}
  \biggr\},
  \end{equation}
\end{widetext}
where $\hat\F^{(k)}(0)$ is the $k$-th derivative of the correlator
$\hat\F(t)$ [Eq.~(\ref{eq:correlator-explicit})] evaluated at
$t=0$. The corresponding maximum decoherence rate with order-one
refocusing (which is determined by {\em reactive\/} non-resonant
processes) can be estimated as
\begin{equation}
  \Gamma_1\sim \max\bigl[J^2\tau,\,\Delta(0)\tau\bigr],
  \label{eq:decoherence-rate-one}
\end{equation}
where $\Delta(t)\equiv \|\hat\F(t)\|$ was defined below
Eq.~(\ref{eq:decoherence-rate-stationary}).  Here the first expression
comes from the first term in
Eq.~(\ref{eq:kernel-non-singular-expanded}) and originates from the
non-compensated evolution due to the system
Hamiltonian~(\ref{eq:ham-system}), while the second term is an
estimate of the leading order of the derivative expansion in
Eq.~(\ref{eq:kernel-non-singular-expanded}).  The presence of the
instantaneous correlator can be interpreted as an effect of nearly
static fluctuations of the coefficients $J_\alpha$ due to the presence
of the bath.  Comparing with the corresponding expression in the
absence of refocusing, we note that already with first-order
refocusing the decoherence rate is reduced as long as the refocusing
is fast enough, $\tau/\tau_0\ll 1$, $J\tau\ll 1$.

With the {\em second-order refocusing\/},
Eq.~(\ref{eq:refocusing-two}), all evolution to quadratic order in $J$
should be compensated.  To demonstrate this cancellation explicitly
for the first term of Eq.~(\ref{eq:kernel-non-singular-expanded}) with
$p=0$, denote
\begin{equation}
\mathcal{M}_{\alpha\beta}\equiv \sum_{m\neq0}
    {[\hat A_{-m}^{\rm tr} J \,J^{\rm tr} \hat
    A_{m}]_{\alpha\beta} \over -i\Omega_{-m}} , \quad \hat\mathcal{M}^{\rm
    tr}=-\hat\mathcal{M}. \label{eq:matrix-m} 
\end{equation}
Then the second-order refocusing condition~(\ref{eq:refocusing-two})
implies 
\begin{equation}
\mathcal{M}_{\alpha\beta}\Sigma_\alpha\Sigma_\beta=
\mathcal{M}_{\alpha\beta}
(B^\gamma_{\alpha\beta}+iC^\gamma_{\alpha\beta})
\Sigma_\gamma=0,\label{eq:refocusing-two-m}
\end{equation}
while the first term in Eq.~(\ref{eq:kernel-non-singular-expanded})
was obtained from the double commutator, 
$$
[\Sigma_\alpha,[\Sigma_\beta,\Sigma_\delta]]= 
\Sigma_\alpha\Sigma_\beta\Sigma_\delta+
\Sigma_\delta\Sigma_\beta\Sigma_\alpha-
\Sigma_\alpha\Sigma_\delta\Sigma_\beta-
\Sigma_\beta\Sigma_\delta\Sigma_\alpha.
$$
Clearly, the first two terms in the corresponding product with
$\mathcal{M}_{\alpha\beta}$ are zero due to the refocusing
condition~(\ref{eq:refocusing-two-m}), while the remaining two terms
 cancel each other due to the antisymmetry of the matrix
$\hat\mathcal{M}$. 

The cancellation works essentially the same way for the terms
involving symmetric matrices, even and odd derivatives of the real and
imaginary parts of the correlator matrix $\hat\F(t)$ respectively,
$\hat\F_1^{(2k)}(0)$ and $\hat\F_2^{(2k+1)}(0)$ (again, we use the
assumption that ``frozen'' bath fluctuations are refocused).  Thus,
under most general conditions, the leading order term in the
derivative expansion will be given by $\hat\F_2(0)$, which gives
\begin{equation}
  \label{eq:decoherence-rate-two}
 \Gamma_2\sim\Delta_2(0)\tau, 
\end{equation}
where $\Delta_2(t)\equiv \|\hat\F_2(t)\|$ is defined in analogy with
$\Delta(t)$ but involves only the imaginary part of the correlator
$\hat\F(t)$.  Formally, this term is of the same order as that
remaining after first-order refocusing,
Eq.~(\ref{eq:decoherence-rate-one}).  We note, however, that this
contribution represents essentially quantum effects; for temperatures
not small compared with the bath cut-off scale,
$\beta\omega_c\lesssim1$, it is expected to be small compared with
$\Delta(t)$.  

In practice, the leading-order contribution to the decoherence rate,
Eq.~(\ref{eq:decoherence-rate-two}), is often suppressed altogether.
Indeed, the entire contribution of $\hat\mathcal{F}_2(t)$ to
Eq.~(\ref{eq:kernel-non-singular-expanded}) is identically zero for
the terms involving a single spin, as the nested
commutator-anticommutator of Pauli matrices vanishes,
$[\sigma_\alpha,\{\sigma_\beta,\sigma_\delta\}]=0$ [the value
  $\beta=0$ is excluded from the implicit summation, cf.\ 
  Eqs.~(\ref{eq:qke-generalized}), (\ref{eq:qke-commuted})]. For more
complicated systems (e.g., involving thermal bath correlated across
several qubits), the matrix $\F(t)$ is expected to be symmetric as
long as the bath is time-reversal invariant, that is, for real-valued
spectral coupling function (\ref{eq:coupling-matrix}), $\hat
F(\omega)=\hat F^*(\omega)$.  In such cases all terms in the
derivative expansion of the second order contribution to decoherence
rate are suppressed, which may result in an exponentially  smaller value of
$\Gamma_2$ for $\tau_0\gg\tau$.
Such a situation where $\hat\F_2(t)\equiv0$ and {\em all\/} orders in
the derivative expansion with $\hat\F_1^{(k)}(0)$ are suppressed are
discussed in Sec.~\ref{sec:examples} [see Figs.~\ref{fig:coef1},
  \ref{fig:coef2}].  Here the second order contribution to the
decoherence rate is seen to be small beyound the numerical precision
already for $\tau_0/\tau\agt1$.

{\bf Initial decoherence.}  The spectrum of the decoherence rates
associated with the modes around a frequency $\Omega_l$, $l\neq 0$ is
determined by the positions of the poles of the corresponding
resolvent, $\biglb(p-i\Omega_l-\hat\Pi_{l,l}^{[l]}(p)\bigrb)^{-1}$, in
the vicinity of $p=i\Omega_l$.  Because of the formal
identity~(\ref{eq:kernel-non-singular-identity}),
$\hat\Pi_{l,l}^{[l]}(p)=\hat\Pi_{0,0}^{[0]}(p-i\Omega_l)$, the
corresponding poles are distributed around $p=i\Omega_l$ in an
identical fashion as those around $p=0$.  As a result, at time moments
commensurate with the refocusing period, $t=\tau$, $2\tau$, $3\tau$,
\ldots, the contributions with all frequencies $\Omega_l$ add
coherently, with the common set of decoherence rates $\{\gamma\}$
whose maximum is determined by Eqs.~(\ref{eq:decoherence-rate-one}),
(\ref{eq:decoherence-rate-two}) depending on the order of the
refocusing sequence\endnote{When reporting preliminary results of this
  work in the conclusion of 
  Ref.~\onlinecite{sengupta-pryadko-ref-2005}, we erroneously stated
  that ``only the dynamics in the slow sector is protected by the
  refocusing.''  While the latter statement should be disregarded, it
  does not reduce the validity of other results of
  Ref.~\onlinecite{sengupta-pryadko-ref-2005}.}.

The decoherence rates $\{\gamma\}$ determine the long-time exponential
fall-off of the refocusing accuracy.  The corresponding prefactor
determines the initial decoherence\cite{falci-2005}; it can be found
as the sum of the (non-singular) matrix elements in
Eqs.~(\ref{eq:series-resum-l}), (\ref{eq:series-resum-zero}).  For $t$
sufficiently small, $\Gamma\,t\ll1$, the correction due to the
decoherence can be neglected, and the net contribution of a sector
with given $l$ can be found as the sum of the residues near
$p=i\Omega_l$.  For example, the total weight associated with the
$l=0$ sector can be obtained from Eq.~(\ref{eq:series-resum-zero}) as
the coefficient in front of $p^{-1}R(0)$ at $\Gamma\ll p\ll \Omega$,
\begin{equation}
  \hat \kappa_0=(\hat1-\hat\pi_0')^{-1}\biggl[\hat 
    1+\!\sum_{m\neq0} {\hat\Pi_{0,m}^{[0]}(0)\over
      -i\Omega_m}\biggr],\;
  \pi_0'\equiv{d\hat\Pi_{0,0}^{[0]}(p)\over 
    dp}\biggr|_{p=0}.
\label{eq:correction-matrix-zero}
\end{equation}
The weight of an $l\neq0$ sector is obtained from
Eq.~(\ref{eq:series-resum-l}), 
\begin{equation}
  \hat \kappa_l={\hat\Pi^{[0]}_{-l,0}(0)\over i\Omega_l}(\hat
  1-\hat\pi_0')^{-1}  
  \biggl[\hat1+\!\sum_{m\neq 0 }{\hat \Pi_{0,m}^{[0]}(0)\over
      -i\Omega_m}\biggr], 
\label{eq:correction-matrix-l}
\end{equation}
and the overall total, $\kappa=\sum_l \kappa_l$, is 
\begin{equation}
  \label{eq:correction-matrix-total}
  \hat\kappa=\biggl[\hat 1+\!\sum_{l\neq0}{\hat\Pi^{[0]}_{-l,0}(0)\over
  i\Omega_l}\biggr](\hat 1-\hat\pi_0')^{-1}  
  \biggl[1+\!\sum_{m\neq 0 }{\hat \Pi_{0,m}^{[0]}(0)\over
      -i\Omega_m}\biggr].
\end{equation}
To quadratic order in powers of the perturbing Hamiltonian (the
accuracy of the employed QKE), and to leading order in the derivative
expansion [cf.\ Eq.~(\ref{eq:kernel-non-singular-expanded})],
\begin{widetext}
\begin{eqnarray}
       \hat\Pi_{0,m}^{[0]}(p)&=& (A_m^{\rm tr} J)_\alpha
     C_{\alpha\delta}^\gamma 
     +\sum_{m'\neq 0}  {[\hat A_{m'}^{\rm tr}\biglb(J J^{\rm
     tr}+\hat\F_1(0)\bigrb) \hat A_{m-m'}]_{\alpha\beta} \over
     p-i\Omega_{m'}} 
  C_{\alpha\alpha'}^\gamma
     C_{\beta\delta}^{\alpha'}, \\ 
     \hat\Pi_{-m,0}^{[0]}(p)&=& (A_m^{\rm tr}
     J)_\alpha   C_{\alpha\delta}^\gamma
     +\sum_{m'\neq 0}  {[\hat A_{m'}^{\rm tr}\biglb(J J^{\rm
     tr}+\hat\F_1(0)\bigrb) \hat 
       A_{m-m'}]_{\alpha \beta} \over
       p+i\Omega_{m-m'}}
  C_{\alpha\alpha'}^\gamma
     C_{\beta\delta}^{\alpha'}.
\end{eqnarray}
\end{widetext}
Performing the expansion to quadratic order in $J$ and linear order in
$\F$ and collecting various terms, we obtain for the overall
coefficient~(\ref{eq:correction-matrix-total}), with the same
accuracy,
\begin{equation}
(\hat\kappa-1)_{\gamma\delta}= [\hat q ^{\rm tr} \hat\F_1(0) \hat
  q]_{\alpha\beta} C_{\alpha\alpha'}^\gamma
C_{\beta\delta}^{\alpha'}, \label{eq:correction-matrix-leading}
\end{equation}
where 
\begin{equation}
\hat q\equiv \lim_{\epsilon\to+0}\int_0^\infty dt
\,e^{-\epsilon t} \hat Q(t)=\sum_{m\neq0} {\hat A_m\over
  -i\Omega_m}.\label{eq:integral-A}
\end{equation}
Note that this expression was derived assuming solely order-one
refocusing, yet the constant coefficients $J^\alpha$ give no
contribution to quadratic order here.  For a generic first- or
second-order refocusing sequence, $0\neq\hat q\sim \tau$, and the
initial decoherence can be estimated as
\begin{equation}
  \label{eq:initial-decoherence-estimated}
  \|\hat\kappa-1\|\sim \Delta(0)\tau^2.
\end{equation}

For sequences which produce time-reversal symmetric evolution, $\hat
Q(t)=\hat Q(-t)$, the Fourier components are real-valued, $\hat
A_m=\hat A_{-m}$, and the sum~(\ref{eq:integral-A}) vanishes
identically.  In such cases the initial decoherence is smaller, and it
is determined by higher derivatives of the bath correlation function,
e.g.,
\begin{equation}
  \label{eq:initial-decoherence-estimated-symmetric}
  \|\hat\kappa-1\|_{\rm symm}\sim |\Delta''(0)|\tau^4
\end{equation}
for the symmetric sequence in Fig.~\ref{fig:coef2}. 

So far we only considered the terms $\sim R(0)$ which depend on the
initial conditions for the density matrix.  The remaining terms in the
r.h.s.\ of Eqs.~(\ref{eq:series-resum-zero}),
(\ref{eq:series-resum-l}) provide an additional source of errors, as
these terms are responsible for establishing the correlations
characteristic for the stationary state at large $t$; they are
required to vanish at $t\to0$.  In real-time the corresponding
contributions come with the prefactors $1-e^{(i\Omega_l-\gamma)t}$,
small at commensurate time moments $t=\tau, 2\tau, \ldots$ because the
decoherence rates $\gamma$ are small.  Additional smallness arises
because the refocusing tends to average out the correlations which
would normally appear as the equilibrium is reached.  Therefore, we
expect these contributions to be quartic, beyound the accuracy of the
present calculation.

\section{Example: single-spin kinetics.}
\label{sec:examples}

In this section we illustrate the derived general expressions on an
example of a single qubit (spin) driven by classical
fluctuating fields.  Specifically, we use Gaussian random fields
$b_\alpha(t)$ along one ($x$) or all three directions, with the
correlators
\begin{equation}
  \langle
  b_\alpha(t)\rangle=0,\quad  \langle
  b_\alpha(t)b_\beta(t')\rangle
  =\delta_{\alpha\beta}b_0^2e^{-t^2/(2\tau_0^2)},
  \label{eq:random-field-correlator}
\end{equation}
where $b_0$ is the r.m.s.\ amplitude of the random field and $\tau_0$
is the correlation time.  The correlated field is generated using the
spectral filter based on fast fourier transformation (FFT) of a
sequence of originally uncorrelated Gaussian random numbers.  As a
result, $b_\alpha(t)$ are actually periodic over the simulation
interval (which is always long compared to $\tau_0$).

The density matrix~(\ref{eq:density-matrix}) is described by the
three-component vector ${\bf R}$, $R^2=1$, whose quantum dynamics is
described by the Bloch equation,
$$
\dot {\bf R}=[{\bf B}(t)\times {\bf R}],
$$
where $B_\alpha(t)=V_\alpha(t)+b_\alpha(t)$ is the net magnetic
field [see Eqs.~(\ref{eq:ham-control}), (\ref{eq:ham-system-bath})].
In terms of the vector ${\bf R}$, the spin evolution in a given
classical field is a rotation; the goal of refocusing is to reduce the
total rotation angle $\phi$.  The average fidelity of the refocusing,
the probability for the qubit to remain in the original state,
averaged over initial conditions, is equal to
$1-(1-\langle\cos\phi\rangle)/3$.

\begin{figure}[htbp]
  \includegraphics[width=\columnwidth]{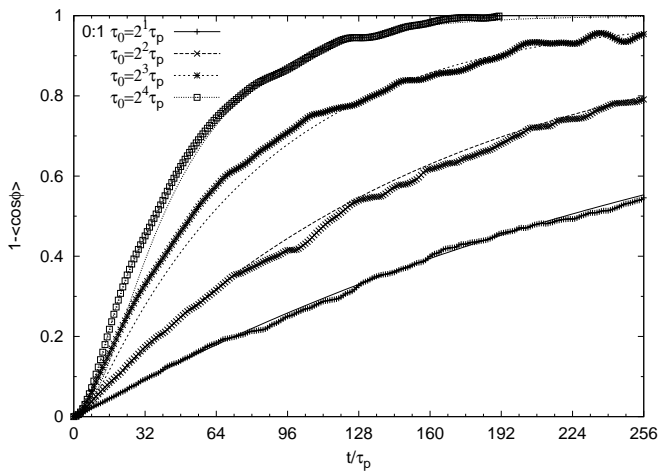}
  \caption{``Refocusing error'' (specifically, three times the deviation
    of the average fidelity for the spin to remain in the initial
    state from one), in the absence of refocusing.  The gaussian
    random field is applied along $x$-axis only.  For different
    curves, it has the same r.m.s.~amplitude $b_0$ but different
    values of the correlation time~$\tau_0$ [see
      Eq.~\ref{eq:random-field-correlator}].  Symbols show the results
    of simulation averaged over 900 samples of random field; lines
    show the corresponding exact results~(\ref{eq:exact-averaging}),
    (\ref{eq:exact-scaling}) which for $t\gg\tau_0$ are also very
    close to the QKE result (not shown).  See text for other
    notations.}
  \label{fig:samp0p1}
\end{figure}

In Fig.~(\ref{fig:samp0p1}) we show the results of time-dependent
simulations for a single spin driven by one-component random field
with four different values of the correlation time $\tau_0$.  We plot
the quantity $1-\langle \cos\phi\rangle$, proportional to the
deviation of the average fidelity for the spin to remain in the same
state from one, as a function of time $t$ in units of $\taup$, a time
scale equal to the interval between consecutive refocusing pulses.
The r.m.s.~amplitude of the random field is the same for all curves
(in fact, everywhere throughout this work), $b_0=0.0355/\taup $.
The numerical data is compared with the exact analytical solution,
\begin{eqnarray}
  \label{eq:exact-averaging}
1-\langle \cos\phi(t)\rangle=1-e^{-\langle\phi^2(t)\rangle/2},\;\,
\phi(t)=\int_0^t b_x(t)\,dt,\;\;\\
{\langle\phi^2\rangle\over2}=b_0^2\tau_0^2(
  \pi^{1/2}x\mathop{\rm erf} x+e^{-x}-1),\;\, x={t\over
    \sqrt2\tau_0}.\quad
  \label{eq:exact-scaling}
\end{eqnarray}
For long-time asymptotics we obtain
\begin{equation}
  \label{eq:exact-averaging-long-time}
  \langle \cos\phi(t)\rangle\to e^{b_0^2\tau_0^2}e^{- \gamma_{\rm
      exact} t},
  \;\, \gamma_{\rm exact}=b_0^2\tau_0
  (\pi/2)^{1/2}.
\end{equation}

To make a connection to the quantum kinetic equation, we notice that
in the simulations we perform the averaging over the classical random
fields, instead of that over the quantum dynamics of the thermal
bath.  As a result, the correlation matrix 
$\F_{\alpha\beta}(t-t')=\langle b_\alpha(t)b_\beta(t')\rangle$ is
explicitly real-valued, $\hat\F_2(t)=0$.  In the absence of the
control fields, the Laplace-transformed
QKE~(\ref{eq:qke-stationary-laplace}) is
\begin{eqnarray}
\label{eq:qke-stationary-laplace-single}
\tilde {\bf R}(p)&=&[p-\hat\Pi(p)]^{-1} {\bf R}(0),\\
\Pi_{\gamma\delta}(p)&=&J_\alpha e^{\alpha\delta\gamma}+\tilde
\F_{\gamma\delta}(p)-\delta_{\gamma\delta} \tilde
\F_{\alpha\alpha}(p).
\label{eq:qke-stationary-laplace-single-kernel}
\end{eqnarray}
With $J_\alpha=0$, and for the random field along the $x$-axis only,
the exponent and the prefactor of the exact long-time asymptotics can
be calculated to quadratic order in the noise amplitude by expanding
the resolvent of Eq.~(\ref{eq:qke-stationary-laplace-single}) around
the point $p=0$, 
$$\tilde {\bf R}(p)\approx [p-p\hat\Pi'(0)-\hat\Pi(0)]^{-1}{\bf R}(0).
$$
This gives in real time [cf. Eq.~(\ref{eq:exact-averaging-long-time})]
$$\langle\cos\phi\rangle\to (1-b_0^2\tau_0^2)^{-1}e^{-\gamma t},\quad 
\gamma={\gamma_{\rm exact}
\over 1-b_0^2\tau_0^2}
$$

The simulations with refocusing were performed using a symmetric
length-$8$ pulse sequence ``8p''
($\X\Y\X\overline\Y\overline\Y\X\Y\X$), as well as a set of
``concatenated'' pulse sequencs, ``4c'' ($\X\Y\overline\X\Y$), ``8c''
($\X\Y\overline\X\Y\overline\X\overline\Y\X\overline\Y$), ``16c''
($\X\Y\overline\X\Y\overline\X\overline\Y\X\overline\Y
\overline\X\Y\X\Y\X\overline\Y\overline\X\overline\Y $), etc, where
$\X$ is a $\pi$-pulse along the $x$-direction, $\overline\X$ is a
negative-$\pi$ pulse, and the longer sequences are obtained
recursively by ramping the signs of the pulses.  This concatenation
procedure is somewhat similar but differs from that used in
Ref.~\onlinecite{khodjasteh-Lidar-2004}.

We used the Gaussian pulses\cite{bauer-gauss}, as well as the first-
and second-order self-refocusing $\pi$-pulses, $S_{L}$ and $Q_{L}$
respectively, designed by the authors
previously\cite{sengupta-pryadko-ref-2005}.  Pulses $S_{L}$, $L=1,2$
are analogous to the first-order Hermitian pulses\cite{warren-herm}
but they were constructed so that the amplitude of the signal (along
with the derivatives up to $2L$-th) turn to zero at the ends of the
interval of the duration $\taup $.  Pulses $Q_{L}$, $L=1,2$ are
similarly designed one-dimensional second-order self-refocusing
pulses.

\begin{figure}[htbp]
  \includegraphics[width=\columnwidth]{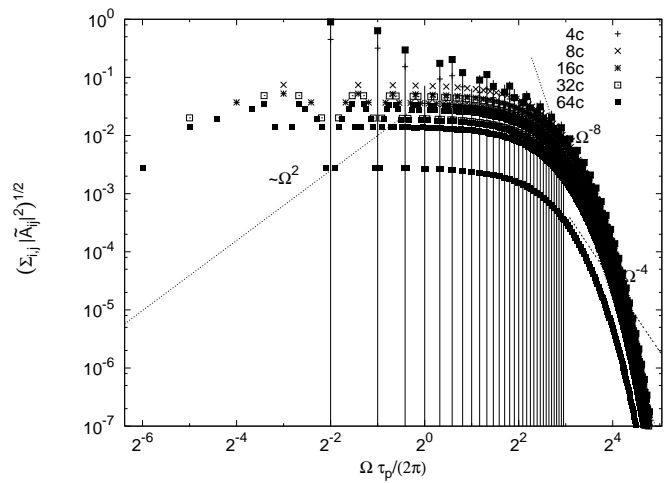}
  \caption{The Frobenius norm of the matrices $\hat A_m$
    [Fourier transform of the evolution matrices $\hat Q(t)$, see
    Eq.~(\ref{eq:rotation-fourier})] with frequencies $\Omega_m=2\pi
    m/\tau_p$ for sequences $n$c, $n=4,8,\ldots,64$ with Gaussian
    $\pi$-pulses.  The width of a pulse is chosen to be $0.05\taup$ so
    that the discontinuity at the ends of the interval is numerically
    negligible, which results in a steep cut-off at
    high frequencies.  Vertical lines mark the spectrum features of
    the parent sequence 4c, which dominate the spectrum of all
    higher-order sequences.  Thin dotted lines guide the eye with the
    slope corresponding to power laws as indicated.}
  \label{fig:spec_c_G005}
\end{figure}

\begin{figure}[htbp]
  \includegraphics[width=\columnwidth]{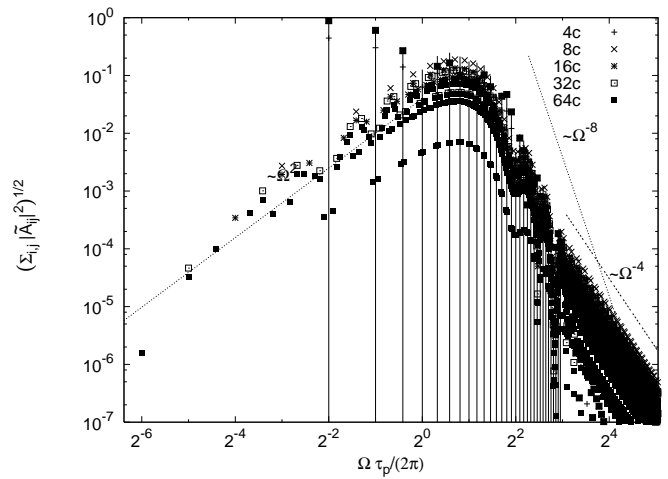}
  \caption{As in Fig.~\ref{fig:spec_c_G005} for first-order
    self-refocusing pulses $S_1$.}
  \label{fig:spec_c_S1}
\end{figure}

\begin{figure}[htbp]
  \includegraphics[width=\columnwidth]{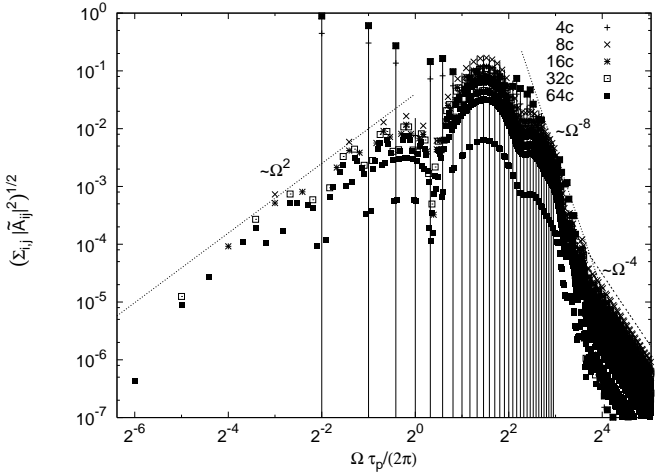}
  \caption{As in Figs.~\ref{fig:spec_c_G005}, \ref{fig:spec_c_S1} with
    second-order pulses $Q_1$.  Note a suppression of the
    low-frequency part of the spectrum compared with lower-order
    pulses.}
  \label{fig:spec_c_Q1}
\end{figure}

\begin{figure}[htbp]
  \includegraphics[width=\columnwidth]{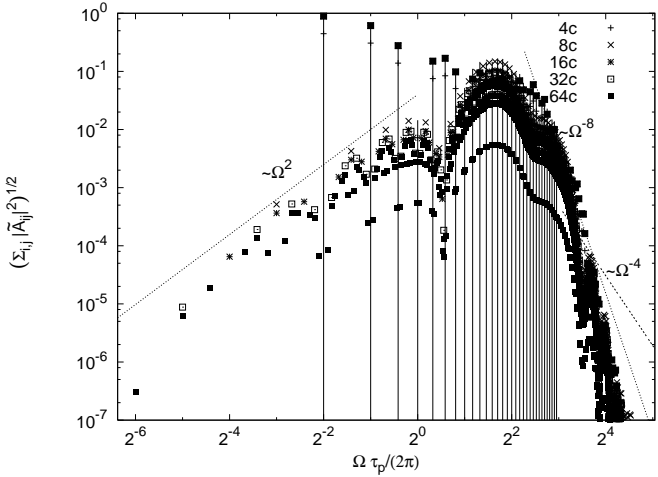}
  \caption{As in Figs.~\ref{fig:spec_c_G005}, \ref{fig:spec_c_S1},
    \ref{fig:spec_c_Q1} for second-order pulses $Q_2$.  These shapes
    vanish at the ends of the interval along with the first three
    derivatives, which suppresses the high-frequency part of the
    spectrum.}
  \label{fig:spec_c_Q2}
\end{figure}

\begin{figure}[htbp]
  \centering
  \includegraphics[width=\columnwidth]{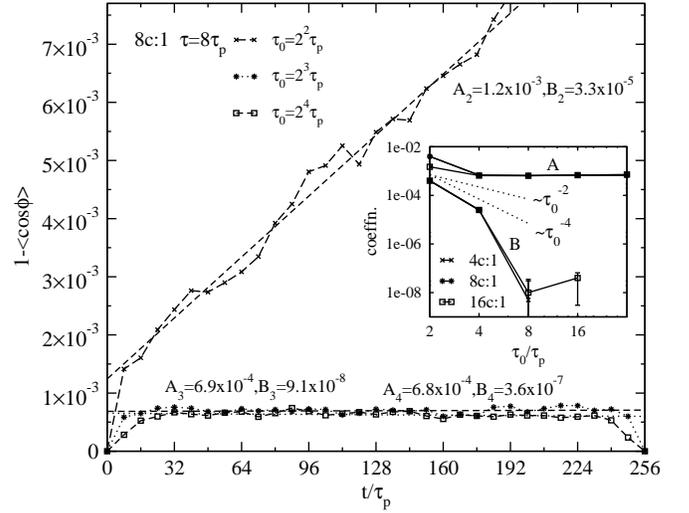}
  \caption{Refocusing error with the fluctuating magnetic field along the
    $x$ axis as in Fig.~\ref{fig:samp0p1}, but now in the presence of
    refocusing sequence 8c with pulses $Q_1$ (order 2, see
    Tab.~\ref{tab:sequences}).  Symbols represent data averaged over
    900 realizations of disorder, dashes are the linear fits
    [Eq.~(\ref{eq:linear-fit-parameters})] for data further than
    $\Delta t=3\tau_0$ from the ends of the interval. Inset shows the
    fit coefficients for sequences 4c, 8c, and 16c as a function of
    the ratio of the correlation time $\tau_0$.  Dotted lines on the
    inset indicate the slope corresponding to the power laws as
    indicated.  The decoherence rate (proportional to the slopes $B$)
    is reduced dramatically for the correlation time $\tau_0$
    exceeding the duration of the sequence, $\tau=n\tau_p$ for
    sequence $n$c.  Yet the refocusing error does not disappear
    altogether because of the initial decoherence (proportional to the
    intercept $A$) which does not vanish with increased noise
    correlation time $\tau_0$ for these non-symmetric sequences [see
    Eq.~(\ref{eq:initial-decoherence-estimated})].  The data on the
    inset also shows that the refocusing accuracy is not improved
    with the longer sequences of order above 2nd, but it also does not
    worsen even for small $\tau_0/\tau$ [the low-frequency
    harmonics $\hat A_m$ are suppressed, see Fig.~\ref{fig:spec_c_Q1}].
    The refocusing errors are strongly suppressed near the end of the
    interval because the fluctuating field used in the calculation is
    periodic with the period $T=256\tau_p$.}
  \label{fig:coef1}
\end{figure}

\begin{figure}[htbp]
  \centering
  \includegraphics[width=\columnwidth]{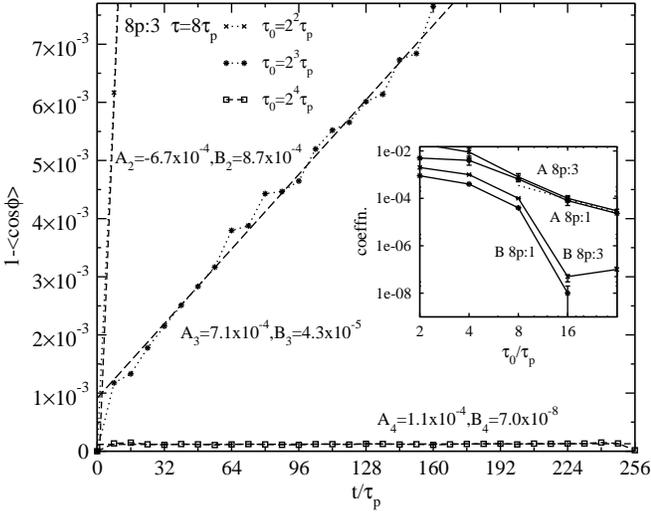}
  \caption{As in Fig.~\ref{fig:coef1} but for the sequence 8p (pulses
    Q$_1$) and for the magnetic field fluctuating in all three
    directions, Eq.~(\ref{eq:phonon-correlator}).  Amplitude $b_0$
    along each direction and other simulation parameters are as in
    Fig.~\ref{fig:coef1}.  Inset shows the linear fit coefficients
    for this seqience with the magnetic field fluctuating in one
    (8p:1) and all three (8p:3) directions.  The effective decoherence
    rate for the fastest fluctuations, $\tau_0=4\tau_p$, is bigger
    then those in the simulations with $n$c sequences.  However, with
    $\tau_0>\tau=8\taup$, the decoherence rate (slope $B$) again goes down
    dramatically, while the initial decoherence also scales down {\em
      quadratically\/} with increasing $\tau_0$, resulting in a
    superior refocusing accuracy. }
  \label{fig:coef2}
\end{figure}

The order of the sequences with the particular pulses and for
different directions of the applied constant field are listed in
Tab.~\ref{tab:sequences}.  The sequence 8p has the duration
$\tau=8\taup $, and so the Fourier expansion of the corresponding
evolution operator $Q(t)$ starts with the frequency
$\Omega=2\pi/(8\taup )$.  Similarly, for sequences 4c, 8c, \ldots, the
Fourier expansion starts with $2\pi/(4\taup )$, $2\pi/(8\taup )$, etc.
However, due to the structure of these sequences, the low-frequency
Fourier coefficients for sequences in the $n$c series with larger $n$
turn out to be very small numerically, and scaling as $\sim \Omega^2$,
as illustrated in Figs.~\ref{fig:spec_c_G005}, \ref{fig:spec_c_S1},
\ref{fig:spec_c_Q1}, \ref{fig:spec_c_Q2}.  As a result, for relatively
fast fluctuations (small $\tau_0$) the long-time refocusing accuracy
for these sequences can be substantially better than that of equal or
shorter ordinary sequences (compare the slopes with $\tau_0=2^2\taup $
in Figs.~\ref{fig:coef1} and \ref{fig:coef2}).  We also note the
suppressed high-frequency tail of the spectra in
Fig.~\ref{fig:spec_c_Q2} which illustrates the advantage of the pulses
$Q_2$ designed\cite{sengupta-pryadko-ref-2005} specifically for
reduced spectral width\cite{borgnat-1996}. 

\begin{table}[htbp]
  \centering
  \begin{tabular}[c]{c||c|c|c|c|c|c|c|c|c|c|}    
    &\multicolumn{3}{c}{{\bf 1}: $B_x\neq0$}\vline
    &\multicolumn{3}{c}{{\bf 2}: $B_z\neq0$}\vline 
    &\multicolumn{3}{c}{{\bf 3}: $B_x,B_y,B_z\neq0$}\vline
    \\   \hline
    seq$\backslash$pulse
    &G&$S_L$&$Q_L$ 
    &G&$S_L$&$Q_L$ 
    &G&$S_L$&$Q_L$ 
    \\  \hline
    4c
    &0&2&2
    &0&1&2
    &0&1&1
\\
    8c
    &2&4&6
    &1&3&5$^*$
    &1&1&1
\\
    16c
    &2&4&6$^*$
    &2&6$^*$&8$^*$
    &1&1&1
\\
    32c
    &2&4$^*$&6$^{**}$
    &4$^*$&8$^{**}$&$\ge10$
    &1&1&1
\\
    64c
    &2&4$^{**}$&6$^{**}$
    &4$^{**}$&$\ge10$$$&$\ge10$
    &1&1&1
\\
    8p
    &1&1&3
    &1&1&3
    &1&1&2
  \end{tabular}
  \caption{Order of refocusing sequences (rows) with different
    pulse shapes (columns), depending on the direction of the applied
    constant field.  The order values listed represent the number of
    cancelled terms in 
    the cumulant expansion~(\ref{eq:cumulants}) of the evolution
    operator with the bath variables replaced by c-numbers.
    ``G'' stands for Gaussian pulses\cite{bauer-gauss,warren-herm}, $S_L$ 
    and $Q_L$, $L=1,2$ are first- and second-order one-dimensional
    self-refocusing pulses respectively with up to $2L-1$ derivatives
    vanishing at the  ends of the
    interval\cite{sengupta-pryadko-ref-2005}.  The superscripts
    ``$\,^*$'' or ``$\,^{**}$'' denote that the first non-vanishing cumulant
    is ``small'' or ``very small'' numerically (smaller by some two
    and four orders of 
    magnitude respectively compared to what is expected from naive
    scaling).  The expansion was done numerically keeping 10 orders in
    the time-dependent perturbation theory as explained in
    Ref.~\onlinecite{sengupta-pryadko-ref-2005}.  See text for  
    definitions of the sequences.}
  \label{tab:sequences}
\end{table}


Figs.~\ref{fig:coef1}, \ref{fig:coef2} show the refocusing error,
$1-\langle\cos\phi\rangle$ with the refocusing pulses present as
described in the captions.  The amplitude of the fluctuating field
$b_\mu(t)$ (along one or three directions) are the same as for data in
Fig.~\ref{fig:samp0p1}, but the vertical scale here is reduced by some
two orders of magnitude.  This totally hides the curvature of the
plots a few correlation times away from the origin, which allows a
linear fit,
\begin{equation}
1-\langle\cos\phi\rangle=A+B
\,t/\tau_0.\label{eq:linear-fit-parameters}
\end{equation}
The coefficients represent the initial decoherence
proportional to the intercepts $A$ with the vertical axis, and the
decoherence rate proportional to the slopes $B$.  We note that
the random field used in the simulations is periodic with the period
$T=256\tau_p$; as a result the overall error is almost entirely
compensated towards the end of the simulation interval.  Respectively,
only the data further than $\Delta t=3\tau_0$ from the ends of the
interval was used in the fits.

The pulse shapes and the fluctuating fields chosen for simulations in
Figs.~\ref{fig:coef1} and \ref{fig:coef2} are such that the pulse
sequences provide at least second order refocusing.  With solely
classical correlations, $\hat\F_2(t)=0$ and $\hat\F_1(t)$ symmetric,
the decoherence rate is expected to go down dramatically with
increasing $\tau_0$.  This is exactly what is seen in Figs.\ 
\ref{fig:coef1}, \ref{fig:coef2}: already at $\tau_0\gtrsim \tau$ the
real-time graphs look almost horizontal and the corresponding slopes
$B$ scale down rapidly with increasing $\tau_0$, so that they become
too small for the numerical precision of the calculation.

While the sequence 8p is explicitly symmetric with respect to the
origin, the sequences $n$c are not.  As a result, with slow
fluctuations, $\tau_0/\tau\gtrsim 1$, the initial decoherence for the
latter sequence tends to a constant value, as can be seen from the
intercepts $A$ in Fig.~\ref{fig:coef1}.  On the other hand, the
intercepts tend to be much 
smaller in Fig.~\ref{fig:coef2}, where the symmetric sequence 8p was
used.  This results in an excellent overall refocusing accuracy.

We have also simulated the spin dynamics under the 4c sequence in the
presence of the fluctuating magnetic field along the $z$ direction,
using Gaussian, S$_1$, and Q$_1$ pulses which provide 0th, 1st, and
2nd order refocusing respectively (not shown).  The decoherence rates
in the three cases are seen as proportional to $\tau_0$, independent
of $\tau_0$, and vanishing rapidly with $\tau_0\gtrsim\tau$, as
expected from the analytic calculations.

\section{Conclusions}

In this work we discussed the kinetics of a quantum system subject to
a pulse-based control fields of arbitrary shape.  We concentrated on
the simplest case of dynamical decoupling, or refocusing, where the
only goal is to cancel any evolution due to intrinsic or extrinsic
couplings.  We solved the problem in the approximation of a
non-Markovian quantum kinetic equation, which limits the accuracy to
quadratic order in powers of the perturbations, but considered the
evolution due to the control fields exactly.  The equations correctly
represent long-time dissipative dynamics.  The corresponding
decoherence rates and the prefactor are evaluated to second order in
powers of the small parameter, the evolution amplitude due to the
perturbation over the period of the refocusing sequence.

We demonstrated that higher order refocusing sequences can be very
effective in cancelling the decohering effects of the couplings to
slow external degrees of freedom.  If in the absence of control the
decoherence rate due to the bath with the characteristic correlation
time $\tau_0$ is $\Gamma_0$
[Eq.~(\ref{eq:decoherence-rate-stationary})], with sufficiently fast
order-one period-$\tau$ refocusing ($\tau\lesssim \tau_0$) the
decoherence rate can be reduced by a factor of $\sim (\tau/\tau_0)$
[Eq.~(\ref{eq:decoherence-rate-one})].  This reduction accounts for
both dissipative and reactive terms and is dominated by the latter, as
long as the driven dynamics is in the spectral gap of the thermal
bath.  With second-order refocusing, the decoherence rate is further
reduced, as it is now determined only by the quantum part of the bath
correlator [Eq.~(\ref{eq:decoherence-rate-two})].  With the bath
coupling time-reversal invariant, additional cancellations are
possible, which may ultimately lead to the decoherence rate (in the
QKE order) smaller than any power of the adiabaticity parameter
$\tau/\tau_0$.

As noted on many occasions in NMR literature, symmetric refocusing
sequences provide for additional cancellations in the evolution
operator and often provide superior refocusing
accuracy\cite{mehring-book}.  Here we show that the symmetry is also
crucial for reducing the initial decoherence, an effective dephasing
which occurs at the beginning of the refocusing sequence.  While
generic first- or higher-order control sequences result in an initial
decoherence proportional to the square of the amplitude of the
fluctuating fields, $\sim \Gamma_0 \tau^2/\tau_0$,
[Eq.~(\ref{eq:initial-decoherence-estimated})], with symmetric
sequences this leading-order contribution is cancelled, which produces
an additional reduction by a power of the adiabaticity parameter
$\tau/\tau_0$.

We illustrated these cancellations by simulations of a single qubit in
the presence of a classical fluctuating magnetic field.  Our
simulations suggest that using non-symmetric refocusing sequences of
order higher 
than two does little to improve the decoherence rate of the controlled
system.  Unlike the formulae which target the scaling, the simulations
also illustrate the actual magnitude of the achieved reduction in
decoherence.

In this work we concentrated on the dynamics of a relatively small
quantum system and ignored the scaling of the decoherence rates with
the size of the system.  For example, the estimate
Eq.~(\ref{eq:decoherence-rate-two}) can be rewritten as an upper bound
on the decoherence rate, in which case it contains an additional
factor of $N$, the number of levels in the controlled system.  Further
studies with specific models of bath coupling are needed to understand
in what cases this scaling with $N$ can be suppressed.  Present
estimates are useful for small, few-qubit systems, or for situations
where thermal bath does not induce long-range correlations between
distant qubits.  We plan to analyze the scaling with the system size
and the range of correlations in the thermal bath in a future
publication.  Another planned extension of this work is to analyze the
quantum kinetics of a system and ways to reduce decoherence during the
operation of a quantum algorithm, without the assumptions that the
control fields are periodic.

\acknowledgments We would like to thank Mark Dykman and Daniel Lidar
for numerous illuminating discussions, and to Kaveh Khodjasteh for
insightful comments on the manuscript.


\end{document}